\documentclass[aps, pra, showpacs, amsmath, twocolumn,superscriptaddress,groupedaddress]{revtex4}
\usepackage{graphics}
\usepackage{graphicx}
\usepackage{subfigure}
\usepackage{float}
\usepackage{longtable}
\usepackage{amssymb}
\usepackage{bm}

\begin{document}

\title{Strong-field approximation for Coulomb explosion of H$_2^+$ by short intense laser pulses}
\author{H. A. Leth}
\affiliation{Lundbeck Foundation Theoretical Center for Quantum System Research,
  Department of Physics and Astronomy, University of  Aarhus, 8000 {\AA}rhus C, Denmark}
\author{L. B. Madsen}
 \affiliation{Lundbeck Foundation Theoretical Center for Quantum System Research,
  Department of Physics and Astronomy, University of  Aarhus, 8000 {\AA}rhus C, Denmark}
  \author{J. F. McCann}
\affiliation{Centre for Theoretical Atomic Molecular and Optical Physics,
School of Mathematics and Physics, Queen's University Belfast,
Belfast BT7 1NN, UK}

\date{\today}

\begin{abstract}

We present a simple quantum mechanical model to describe Coulomb
explosion of H$_2^+$ by short, intense, infrared  laser pulses. The
model is based on the length gauge version of the molecular
strong-field approximation and is valid for pulses shorter than 50
fs where the process of dissociation prior to ionization is
negligible. The results are compared with recent experimental
results for the proton energy spectrum [I. Ben-Itzhak et al., Phys.
Rev. Lett. $\textbf{95}$, 073002 (2005), B. D. Esry et al., Phys.
Rev. Lett. $\textbf{97}$, 013003 (2006)]. The predictions of the
model reproduce the profile of the spectrum although the peak energy
is slightly lower than the observations. For comparison, we also
present results obtained by two different tunneling models for this
process.
\end{abstract}


\maketitle

\section{Introduction}

Ultrashort, highly-intense, laser pulses at infrared wavelengths are
currently being used to study molecular dynamics under extreme
conditions. These systems are commonly  based upon a Ti:Sapphire
laser, tunable around 790 nm, which can reach peak intensities in
the range 10$^{13}$-10$^{17}$ W/cm$^{2}$,  with pulse lengths
shorter than 50 fs. Consequently energy  can be deposited on
time-scales shorter than the fastest molecular vibration with field
strengths comparable to the molecular bond. When molecules are
exposed to such an environment, the response is highly-nonlinear and
generally leads to multiple dissociative ionization and high-order
optical scattering. Indeed the multiple fragmentation process for a
heavy polyatomic multi-electron system prevents a detailed analysis
of the energy transfer, simply due to the proliferation of fragments
produced. Small molecular systems, on the other hand, have simpler
structure and fewer relaxation channels. Moreover, the fundamental
molecules such as  H$_2$ or D$_2$,  have an intrinsic value owing to
their fast vibration. Since a 790 nm pulse has a cycle period of 2.6
fs, and perhaps 10-50 fs duration, the vibration provides an
additional internal clock that records the response of the
electronic excitation during the passage of the pulse. Consequently,
there has been extensive study of the interaction of intense field
dissociative ionization of H$_2$  an H$_2^+$ with progressively
shorter and more intense pulses and, in particular, using the
ejected proton energy spectrum as an indicator of the electron
response and a diagnostic of the pulse itself. For reviews of
progress in this field one can consult, for example,
Refs.~\cite{review1995,belfast1999,belgien2004}.

The effect  of the driving pulse on the rates of dissociation and
ionization as well as the energy of the fragments has been discussed
in Refs.~\cite{kansas2005,california1996}. It has been noted that the
kinetic energy release (KER) spectra indicated a decrease in energy
of the ionization products when the pulse duration was increased
~\cite{garching2004a}. For a, comparatively long, 90 fs pulse,  the
observation of a proton energy  peak near 2 eV would correspond to
an internuclear separation $ R \sim  7 $ a$_0$ at the instant of
ionization. Given the equilibrium bond length of H$_2^+$  is $R=2$
a$_0$, this suggests  dissociation occurring prior to ionization.
The increased probability  of ionization with increasing $R$ can be
understood by the mechanism of resonantly-enhanced multiphoton
ionization via the antibonding 2p$\sigma_u^+$  state
~\cite{garching2002,garching2004} when the resonances are tuned
'from the red' as the bond expands. Conversely, at very high
intensities the ionization process can be characterized as a
tunneling transition. The associated shape resonances of the
molecular ion produce the effects known as charge-resonance enhanced
ionization  ~\cite{canada1995} and dynamic tunneling ionization
~\cite{belfast2003}. These ionization mechanisms, diabatic
(multiphoton) and adiabatic (tunneling) ionization, can be studied
indirectly via the proton emission spectrum.

Perhaps the most interesting aspect of current experiments  is that
the laser intensities are high enough to explore the  adiabatic and
diabatic regimes, and also pulse durations are short enough to probe
the vibrational motion. The study of bond-stretching during the
ionization process has been made possible by enhanced control over
pulse duration \cite{kansas2005,kansas2005a} and through  pump-probe
techniques \cite{garching2006,frankrig2006,belfast2007}. In
Ref.~\cite{frankrig2006} it was shown that if H$_2$ was subjected to
two ultra-short pulses, the relative yield of low-energy protons
raised at a probe delay of 24 fs. Similar evidence for dissociation
prior to ionization was reported in Ref.~\cite{garching2002} using
single 100 fs pulses on a H$_2^+$ target. In this case, the
KER-spectrum was explained by one- or two-photon absorption to the
dissociation channel followed by ionization at a separation of 12
$a_0$ giving a structure reflecting the initial distribution of
vibrational states. All of these experiments resulted in a maximum
kinetic energy of the protons in the order of 4 eV.

To minimize the effect of dissociation and thereby be able to
observe protons with higher energies, shorter pulses  are needed.
Experiments using 45 fs pulses showed KER-spectra in the range of
2-10 eV when the intensity exceeded 1.7$\times 10^{14}$ W/cm$^{2}$
\cite{kansas2005,kansas2005a,kansas2006}. At lower intensities the
dissociation channel dominated completely and no Coulomb explosion
was observed. The structure of these spectra was very recently
qualitatively explained by associating the critical bond lengths
with  the Floquet (dressed) state~\cite{kansas2006}  crossing with
the ionic ($1/R$) curve. The underlying physical reason is that the
density of states is very high at threshold and there will be a
resonant coupling to a quasi-continuum. The strength and width of
the peaks are partly restricted by the number of photons involved,
and whether or not the pulse is long enough for the molecular
expansion to occur, and so reach the crossing before the pulse ends.
Using this model, and fitting  a 12-parameter function,  gives  very
good agreement with the observations.

In this work we present a simple quantum mechanical model designed
to predict the high-energy part of the KER-spectrum of dissociative
ionization of H$_2^+$. We use the strong-field approximation (SFA)
within the length gauge ~\cite{aarhus2004,aarhus2005a}. That is we
calculate transition rates from  a molecular state, in which the
laser field is neglected,  to a state in which the laser field is
accounted for to all orders, but some approximations are made for
the three-body continuum. The calculations are simplified by
assuming the Born-Oppenheimer separation of motions, and the
Franck-Condon principle is invoked for the nuclear motion. These
approximations allow us to determine the dissociative ionization
{\em rates} to different channels in a simple way. The pulse is
presented by a Fourier expansion so that the temporal intensity
variation is taken into account. This averaging is important due to
the ponderomotive shift of the electron which may cause channel
closings when the intensity is raised.

Following this procedure, and integrating over the electron
spectrum, the proton energy distribution can be obtained. Comparing
with recent experiments \cite{kansas2005,kansas2006}, we find the
model predictions to be in very good qualitative agreement for the
shape of the distribution. However, our calculations predict the
peak of the proton spectrum at slightly lower energies than that
observed in experiment.
%The discrepancy could be explained by the
%neglect of higher-order processes.
Finally, for completion, we compare the predictions of our model
with tunneling theory calculations. Atomic units
($|e|=\hbar=m_e=a_0=1$) are used throughout unless indicated
otherwise.

\section{Model}

We consider the transition from an initial field-free state to a
state in which the electron is only affected by the laser light, and
the two protons are only subject to their mutual Coulomb repulsion.
There are three essential elements of the model. Firstly, the
initial state, in which the vibrational population distribution
plays a key role. Secondly, the overlap of these vibrational states
with the final Coulomb states of the proton pair is of great
importance in modulating the proton spectrum. Thirdly, and most
importantly, the coupling to the continuum, the nonpertubative
photoionization rate, as a function of bond length and laser
intensity will determine the range of proton energies. All three
factors are intrinsically linked.

Consider a monochromatic plane-wave component of the light field,
with linear polarization,  $\hat{z}$, and angular frequency $\omega$
so that the vector potential in the dipole-approximation is
\begin{eqnarray}
    \bm{A}(t)=A_0\hat{z} f(t)\cos \omega t  \ , \label{eq:Afelt}
\end{eqnarray}
where $A_0$ is the field amplitude, and the pulse shape is described
by the factor, $0 \leq f(t) \leq 1$, which we take to be a Gaussian
profile.

\subsection{Molecular states}

The process of formation of H$_2^+$, requires ionization of the
neutral species. The molecular ion H$_2^+$ has only a single bound
electronic state (1$s\sigma_g$) that is exactly known. The nuclear
relaxation that follows this primary ionization is not so well
defined and remains a source of uncertainty and investigation
~\cite{belgien2004}. Nonetheless, to a very good approximation, the
rotational degrees of freedom can be considered as frozen since its
characteristic timescale of $\sim 170$ fs is much longer than the
pulse duration. However, this results in an ensemble of vibrational
modes, as observed in experiment ~\cite{belfast2005}. As usual, the
$z$-axis of the laboratory reference frame is defined by the
polarization vector, while the $z$-axis of the molecular
(body-fixed) frame is defined by the internuclear axis. The notation
for the coordinates is that $\bm{R}$ denotes the internuclear vector
and the electron coordinate with respect to the internuclear
midpoint, is denoted by $\bm{r}$. In the following presentation, for
consistency, the laboratory reference frame is employed. If we let
$\nu$ denote the vibrational quantum number, then the eigenstates of
the initial ensemble can be expressed as:
\begin{eqnarray}
    \Psi_{i\nu}(\bm{R},\bm{r},t)=\phi_{i}(\bm{r},\bm{R})\chi_{i\nu}(R)
    e^{-iE_{i\nu}t},
\end{eqnarray}
where $\phi_{i}(\bm{r},\bm{R})$ is the electronic wave function,
with  energy $\varepsilon_{i}(R)$, and $\chi_{i\nu}(R)$ is the
vibrational eigenfunction with eigenvalue $E_{Ni\nu}$. The total
energy is then, $E_{i\nu}=\varepsilon_{i}(R_0)+E_{Ni\nu}$. Recall
that the electronic function has a $\sigma_g^+$ symmetry, and the
transformation of
 $\phi_{i}(\bm{r},\bm{R})$  from laboratory to molecule frame is effected by
the Wigner $D$ rotation matrix~\cite{wigner}.

In the  final state, given the large separation between the nuclei
and the electron at the time of ionization, we suppose that the
influence of the protons on the electron is negligible compared to
that of the external field \cite{aarhus2005a}. This is consistent
with the asymptotic ($t \rightarrow +\infty$) limit of the system as
a (decoupled) product state of an outgoing Volkov wave $\phi_{f}$
(electron in the electromagnetic field) and a Coulomb wave
$\chi_{f}$ for the  proton motion:
\begin{eqnarray}
    \Psi_{f}(R,\bm{r},t)=\phi_{f}(\bm{r},t)\chi_{f}(R,t).\label{sluttilstand}
\end{eqnarray}

For an electron in a laser field described by (\ref{eq:Afelt}), the
length-gauge Hamiltonian is given by
\begin{align}
    H_{f}^{elec}&=\frac{1}{2}p^{2}+\bm{r}\cdot\bm{F},\label{hamilton}
\end{align}
where $\bm{p}$ is the canonical momentum and, ${\bm F}= -\partial_t
{\bm A}$, represents the electric field. The Volkov states form a
complete set of solutions to the equation
 $  H_{f}^{elec}\phi_{f}(\bm{r},t)=i\partial_t\phi_{f}(\bm{r},t)$
 and can be written:
\begin{eqnarray}
    \phi_{f}(\bm{r},t)=\exp\left[i(\bm{q}+\bm{A}(t))\cdot
    \bm{r}-i\int_{-\infty}^t\frac{(\bm{q}+\bm{A}(t'))^{2}}{2}dt'\right],
\end{eqnarray}
where $\bm{q}$ is the kinematic momentum corresponding to a drift
energy $q^2/2$, and we denote the ponderomotive energy as $U_p=
A_0^2/4$.

Since vibrational energies are much larger than rotational energies,
and Coriolis coupling can be neglected at these energies, we make
the usual assumption that the proton ejection occurs along the
internuclear axis direction without rotation: the axial recoil
approximation. Thus, the nuclear motion is governed by the
one-dimensional Coulomb repulsion:
\begin{eqnarray}
    H^{nucl}_{f}=-\frac{1}{2\mu}\frac{\partial^{2}}{\partial
    R^{2}}+\frac{1}{R},
\end{eqnarray}
where $\mu= \frac{1}{2}m_p$ is the reduced mass. The corresponding
eigenfunction, with energy  $E_{Nf}$, and wavenumber,
$k_f=\sqrt{2\mu E_{Nf}}$, has the form,
\begin{eqnarray}
    \chi_{f}(R)&=\sqrt{\frac{2\mu}{\pi
    k_f}}F_{0}(\frac{\mu}{k_f};k_fR),\label{kernebolge}
\end{eqnarray}
where
\begin{align}
F_{0}(\frac{\mu}{k_f};k_fR)= &\exp \left(
-\frac{\pi}{2}\frac{\mu}{k_f}+ik_fR \right) \left|\Gamma \left(
1+i\frac{\gamma}{k_f}\right) \right|k_fR \\
         &  \times _1F_1(1+i\frac{\mu}{k_f};2;-2ik_fR), \nonumber
\end{align}
and $_1F_1$ is the confluent hypergeometric series. Since the Volkov
wave represents all orders of the field amplitude, the final state
is a coherent sum of the full spectrum of photoelectron harmonics,
including the angular distribution. Since our primary interest here
is the comparison with experiment for the proton energy spectrum, we
do not present results for the photoelectron differential yields.
Instead we must integrate over these degrees of freedom.
Furthermore, the initial vibrational state is a  mixed state and the
orientation of the molecule is random. In all, this amounts to
integrating over four continuous variables and summing over two
discrete variables, in addition to the  matrix element calculation.
However, since numerical quadrature is inherently a set of
independent calculations, these calculations can readily be
performed on a parallel computer.

\subsection{Transition rates}

To derive the expression for the transition amplitude, we follow the
procedure of Ref. \cite{sydney1997}, and generalize to the case of
an incoherent mixture of initial vibrational states each populated
with probability $P_{\nu}$, defined by the Franck-Condon factors. In
this way we obtain the rate $w$ for a transition into the final
state $\Psi_{\nu}$ of Eq. (\ref{sluttilstand}).
\begin{equation}\label{delta}
    w=\sum_{\nu}P_{\nu}\sum_{n=n_0}^{\infty}2\pi
    \delta(E_{f}-E_{i\nu}-n\omega)\left|A_{\nu
    n}\right|^{2},
\end{equation}
where
\begin{eqnarray}
    A_{\nu n}=\frac{1}{T}\int_{0}^{T}\left\langle \Psi_{f}
    \left|\bm{r}\cdot\bm{F}\right|\Psi_{i\nu}\right\rangle dt, \label{nummer}
\end{eqnarray}
and the minimum number of photons absorbed, $n_0$, is determined by
energy conservation. For a given number of absorbed photons, $n$,
the electron momentum is defined as:
\begin{eqnarray}
q=\sqrt{2(E_{i\nu}+n\omega-U_{p}-E_{Nf})}.
\end{eqnarray}

To obtain the dissociative ionization rate, we multiply by the
density of states per unit energy, per unit solid angle. Using the
normalization convention defined above, the appropriate factor is
$(2\pi)^{-3}q d^{2}\hat{\bm{q}}$. So that we have:
\begin{align}
\label{eq:totrate} \frac{dw}{dE_{Nf}} &=\sum_{\nu}P_{\nu}\int
d\hat{\bm{q}} \sum_{n=n_{0}}^{\infty} \frac{
q}{(2\pi)^2}\left|A_{\nu n}\right|^{2},
\end{align}
where $d\hat{\bm{q}}$ defines the direction of the outgoing
electron.

The calculation of $A_{\nu n}(\bm{q})$ is significantly simplified
in the Franck-Condon approximation, where it is assumed that the
electronic transition appears almost instantaneously compared to
changes in the nuclear position. That is, we can make the
integration over the electron coordinate independent of the nuclear
coordinate by replacing $R$ by some fixed value $R_{0}$,
\begin{align}
\label{eq:A}
   A_{\nu n}=&S_{fi}\frac{1}{T}\int_{0}^{T}D_{el}(\bm{R}_0,t)
e^{i(E_{Nf}-E_{i\nu})t}dt,
\end{align}
where the Franck-Condon factor $S_{fi}$ and the electronic matrix
element $D_{el}(\bm{R}_0,t)$ is given by
\begin{align}
   S_{fi}=&\int_{0}^{\infty}\chi^{\ast}_{f}(R)\chi_{i\nu}(R)dR,\label{franckcondon}\\
   D_{el}(\bm{R}_0,t)=&\int\phi^{\ast}_{f}(\bm{r},t)\bm{r}\cdot\bm{F}\phi_{i}(\bm{r},\bm{R}_{0})d^{3}\bm{r}.
\end{align}
We have tested the validity of the Franck-Condon approximation and
found it to be accurate, especially for small $\nu$.

The evaluation of the matrix element is conveniently carried out in
spherical coordinates~\cite{aarhus2005a}. The first step is to
rewrite the expression as
\begin{align}
  D_{el}(\bm{R}_{0},t)=&\left(E_{i\nu}-E_{Nf}-\frac{(\bm{q}+\bm{A})^{2}}{2}\right)\\
     &\times\exp\left[i\int_{-\infty}^t\frac{(\bm{q}+\bm{A}(t'))^{2}}{2}dt'\right]\tilde{\phi}_{i\nu}(\bm{q}+\bm{A},\bm{R_0})\nonumber,
\end{align}
where we have used
\begin{align}
    -i\frac{\partial\phi_{f}^{\ast}}{\partial t}&=\left[\frac{p^{2}}{2}+\bm{r}\cdot\bm{F}\right]\phi_{f}^{\ast},
\end{align}
and $\tilde{\phi}_{i}$ denotes the Fourier transform of the
electronic wave function,
\begin{eqnarray}
    \tilde{\phi}_{i}(\bm{k},\bm{R_0})=\int\exp\left[-i\bm{k}\cdot\bm{r}\right]\phi_{i}(\bm{r},\bm{R_0})d^3\bm{r}.\label{integral}
\end{eqnarray}
Here $\bm k = \bm q + \bm A$. In the length gauge formulation of the
SFA, the transition amplitude only depends on the asymptotic form of
the coordinate space initial electronic state
\cite{aarhus2004,aarhus2005a,aarhusextra,sydney1997}. In the
laboratory fixed frame this electronic wave function for nuclear
orientation ${\bm R_0}$ reads
\begin{align}
\label{eq:wf}
    \phi_{i}(\bm{r},\bm{R_0}) =\ r^{(\frac{2}{\kappa}-1)}
    \exp\left(-\kappa r\right)\sum_{l}\sum_{m}C_{l0}
    D_{m0}^{(l)}(\hat{\bm{R_0}})Y_{lm}(\hat{r}),
\end{align}
with $C_{l0}$ asymptotic expansion coefficients~\cite{kansas2002},
and
\begin{align}
    \kappa=\sqrt{2\left(\frac1{R_0}-\epsilon_i\right)},
\end{align}
where $\epsilon_i$ is the eigenvalue of the electronic Hamiltonian
including the nuclear repulsion. In Eq.~\eqref{eq:wf}
$D_{m0}^{(l)}(\hat{\bm{R_0}})$ is the Wigner rotation function that
effectuates the transformation from the molecular to the laboratory
fixed frame.

Finally the electronic matrix element can be found as
\begin{align}
    D_{el}(\bm{R}_{0},t)=&\left(E_{i\nu}-E_{Nf}-\frac{(\bm{q}+\bm{A})^{2}}{2}\right)\nonumber\\
    &\times\exp\left[i\int_{- \infty}^t\frac{(\bm{q}+\bm{A}(t'))^{2}}{2}dt'\right]\nonumber\\
    &\times4\pi \sum_{l}\sum_{m}(-i)^{l}C_{l0}D^{(l)}_{m0}(\hat{\bm{R_0}})\ Y_{lm}(\hat{k})\nonumber\\
    &\times\int_0^{\infty} j_{l}(kr)r^{(\frac{2}{\kappa}-1)} \exp \left(-\kappa r\right)r^{2}dr.
\end{align}
The radial integral has a closed analytic form in terms of Gauss's
hypergeometric function~\cite{Gradshteyn}. The time-integration is
performed numerically, along with the sum over the number of
photons. In the experimental spectra, the protons resulting from
dissociative ionization are collected over a range of ejection
angles, with respect to the polarization direction. Averaging over
the different orientations of the molecular axis is equivalent,
within the axial-recoil model, to summing over the ejected proton
directions described by the rotation matrices. This finally reduces
to an energy-differential electronic rate, $\Gamma_{\nu}(E_{Nf})$,
\begin{align}
    \Gamma_{\nu}(E_{Nf})=&q\int d\hat{\bm{q}}\sum _{n=n_0}^\infty \frac{1}{(2\pi)^2}
    \nonumber\\
    &\times\left|\frac1T\int_0^TD_{el}(R_0,t)
    e^{i(E_{Nf}-E_{i\nu})t}dt\right|^2.
\end{align}
To obtain the total rate we use Eqs.
\eqref{eq:totrate}-\eqref{eq:A}, i.e., we multiply by the Franck
Condon factor
\begin{align}
    \frac{dw_\nu}{dE_{Nf}}=|S_{fi\nu}(E_{Nf})|^2
    \Gamma_\nu(E_{Nf})
        \label{rate}
\end{align}
and sum over the different initial states
\begin{align}
    \frac{dw}{dE_{Nf}}=\sum_\nu P_\nu\frac{dw_\nu}{dE_{Nf}}.
\end{align}
The values of  $P_\nu$ are determined by the formation mechanism of
the ion. In the context of recent experiments
\cite{kansas2005,kansas2006} it is reasonable to assume this to be a
Franck-Condon distribution.

Finally, to compare with experimental data, we average over the
pulse profile, $f(t)$ [Eq. \eqref{eq:Afelt}], which is taken to have
a Gaussian profile with FWHM 45 fs. Under the assumption that the
variation in the pule envelope is slow compared with the optical
period, the definition of ionization rate for fixed $A_0$  is still
valid. The ionization process can be significant for intense pulses,
and thus we should allow for depletion of the molecular state. To
model this process, the total ionization probability is found by
integrating this rate over time
\begin{align}
    P_I(E_{Nf})=\sum_\nu\int_{-\infty}^{\infty}\frac{dw_\nu(t)}{dE_{Nf}}N_\nu(t)dt.
\end{align}
Here $N_\nu(t)$ denotes the population in a given vibrational state
$\nu$. This population can be found from the rate equation:
\begin{align}
    \frac{dN_\nu(t)}{dt}=-\Gamma_{tot}^\nu N_\nu(t),
\end{align}
with the boundary condition $N_\nu(-\infty)=P_\nu$ and
$\Gamma_{tot}^\nu$ denoting the total rate of ionization from the
vibrational state $\nu$. By integrating the differential rate over
all final states this rate is found as
\begin{align}
    \Gamma_{tot}^\nu=\int_0^{\infty} \frac{dw_\nu(t)}{dE_{Nf}}dE_{Nf}.
\end{align}

\section{Results}

\begin{figure}
    \centering
    \includegraphics[width=6.8cm,angle=-90]{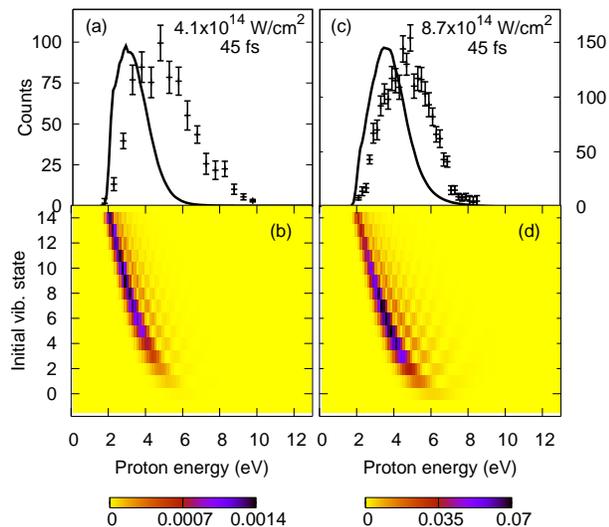}
    \caption{\small{The predicted KER-spectrum for pulses characterized by a wavelength of 790 nm, a duration of 45 fs (FWHM) and a peak
    intensity of $4.1\times10^{14}$ W/cm$^2$ (a) and $8.7\times10^{14}$ W/cm$^2$ (c).
    We also plot the experimental results obtained with these pulses in \cite{kansas2005} (panel (a)) and
    \cite{kansas2006} (panel (c)), respectively. Panels (b) and (d) show the contributions to the
    overall spectrum from the different initial vibrational states. The darker the shading the larger the contribution to a given final proton
    energy from a particular initial vibrational level. }}
    \label{SFA}
\end{figure}

In panels (a) and (c) of Fig.~\ref{SFA} we show the predicted
KER-spectra for Coulomb explosion of H$_2^+$ for two pulses used in
recent experiments \cite{kansas2005,kansas2006}. The spectrum
corresponding to a pulse duration of 45 fs and a maximum intensity
of $4.1\times10^{14}$ W/cm$^2$ peaks near 3 eV, while the spectrum
is moved to slightly higher energies for pulses with maximum
intensity of $8.7\times10^{14}$ W/cm$^2$. Along with the theoretical
predictions, experimental results are also given
\cite{kansas2005,kansas2006}. The qualitative shape of the
distributions is in good agreement  in Fig.~\ref{SFA} (c). However,
the experimental results at the lower intensity, Fig.~\ref{SFA} (a),
show a much broader distribution, with a significant amount of fast
protons (energies above 6 eV) not predicted by theory.

To explain the structure of the theoretical spectra we study the
contributions from the different initial vibrational states. Since
each component contributes incoherently,  they can be studied in
isolation, and we plot in Figs.~\ref{SFA}(b) and (d) these
contributions to the overall spectrum. Here we see that even though
the molecular ions are predominantly  in low vibrational states, the
bulk of the Coulomb explosion yield is from highly vibrationally
excited states. For pulses of peak intensities near
$4.1\times10^{14}$ W/cm$^2$ ionization from the vibrational states
$\nu=6-11$ is favored, while the $\nu=4,5$ and 6 states dominate for
pulses of peak intensities near $8.7\times10^{14}$ W/cm$^2$. It is
this shift in origin of the protons that causes the shift in energy
of the proton spectrum. Highly vibrationally excited molecular ions
contribute to the low energy part of the spectrum, while low-excited
molecular ions result in protons of a higher energy.

\begin{figure}
    \centering
    \includegraphics[width=8.0cm,angle=-90]{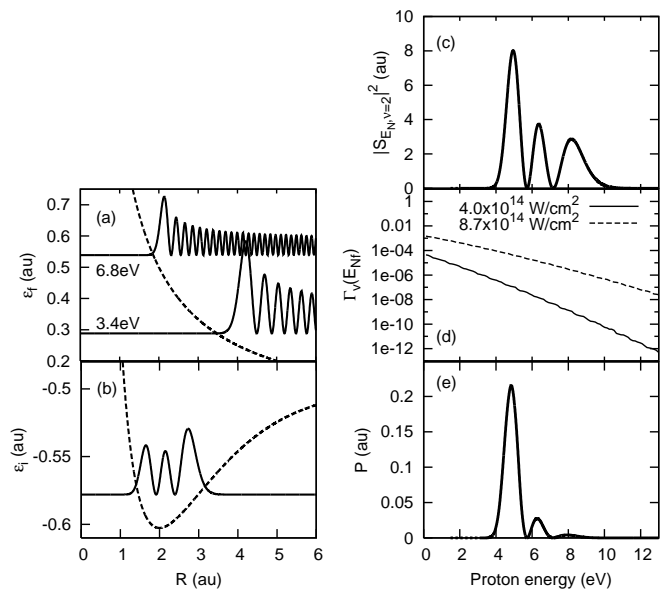}
    \caption{\small{Calculations showing the result for ions in a vibrational excited initial state with $\nu=2$.
    The initial and final wave functions squared are shown in panels (a) and (b). The energies written in panel (a) indicate the proton energy that the particular states correspond to (i.e., one half of the energy of the state).
    Panel (c) shows the Franck-Condon overlap between the initial vibrational state and the final Coulomb wave for the internuclear coordinate. Panel (d) shows the electronic ionization rate for two different intensities
    and panel (e) shows the ionization probability using pulses of 45 fs duration (FWHM), 790 nm wavelength and a peak intensity of 8.7$\times 10^{14}$W/cm$^2$, again for $\nu$=2.  }}
    \label{tredele}
\end{figure}

To see how the structure of the contributions comes about, we
examine the initial and final nuclear wave functions. In panel (a)
of Fig.~2 the probability densities of two energy-normalized
continuum states are presented, along with the density of the
$\nu=2$ state in panel (b). The well known structure of $\nu+1$
peaks with large probabilities near the classical turning points is
seen here. Since we have a continuum of final states which are all
peaking at different internuclear separations, this initial
structure is reflected in the Franck-Condon overlap
$S_{fi\nu}(E_{Nf})$ shown in Fig.~2(c). Remember that large
internuclear separations $R$ correspond to low energies. In
Fig.~2(d) we show the electronic transition rates for two different
intensities. The rates decay, roughly exponentially, with increasing
ejected proton energy with an attenuation more pronounced for the
higher intensity. When multiplying the Franck-Condon factors and the
electronic transition rates and integrating over time we obtain the
final result for the $\nu=2$ case in Fig.~2(e). The influence of the
Franck-Condon factor is still apparent but the attenuation produced
by the decrease in electronic transition rate suppresses all but the
low energies in the proton spectrum. This is reflected in Figs.~1(b)
and 1(d), which isolate the contribution of the different
vibrational states weighted by their populations. The dominant peak
in the contribution is moved towards lower energies as the molecular
ion gets vibrational excited, reflecting the structure of the
nuclear wave function showing a classical turning point at a larger
nuclear separation.

Comparing Fig.~1(b) and (d) shows that at the higher intensity each
vibrational component is broader and the combined spectrum extends
to higher energies, which is also expected due to the less rapid
decrease of the electronic ionization rate at this intensity. In
addition the largest contributions now come from less excited
vibrational states. The combination of  increase in the high-order
multi-photon ionization rate and the relatively larger population in
the lower $\nu$ states compared to the higher $\nu$ states leads to
this enhancement of faster protons.

Overall we see a movement of the predicted spectrum towards higher
energies as the intensity is raised, both due to the change in the
contributions from the different vibrational states and due to the
favoring of low excited vibrational states. This tendency is
reflected in the measurements, but the resemblance is far from
perfect. The lack of agreement may be attributed to several factors.
Clearly,  a limitation of our model is the calculation of ionization
using the strong-field model which is relevant to adiabatic
quasi-tunneling. If resonant-enhanced ionization at smaller bond
lengths were significant, this would produce faster fragment
protons. Also rescattering effects are ignored, which might as well
give protons of high energy.

\subsection{Results using tunneling}

\begin{figure}
    \centering
    \includegraphics[width=6.5cm,angle=-90]{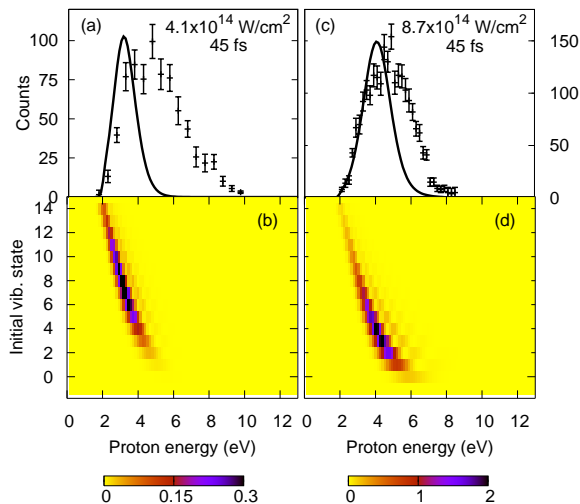}
    \caption{\small{As Fig.~1, but now using  the tunneling theory of
    Ref.~\cite{kansas2002}.}}
    \label{tun1}
\end{figure}
\begin{figure}
    \centering
    \includegraphics[width=6.5cm,angle=-90]{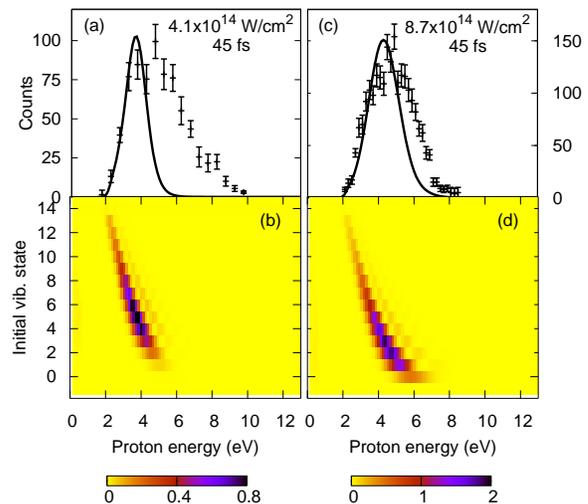}
    \caption{\small{As Fig.~1, but now using the tunneling theory of
    Ref.~\cite{Urbain2004}. }}
    \label{tun2}
\end{figure}

To complete the discussion of Coulomb explosion of H$_2^+$ the
results using a simple tunneling theory are given. Here the
frequency dependence of the field is neglected, and the laser is
treated as a slowly-varying (quasi-static) electric field that
produces tunneling ionization. Strictly speaking, this static model
is only valid when the tunneling time is much shorter than the
optical cycle as expressed by $\gamma_K \ll 1$, where
$\gamma_K=\sqrt{\frac{I_P}{2U_P}}$ is the Keldysh parameter.  For
$I=4.1 \times 10^{14} {\rm W} \ {\rm cm}^{-2}$, $\gamma_K \sim 0.8$,
while for $I=8.7 \times 10^{14} {\rm W} \ {\rm cm}^{-2}$, $\gamma_K
\sim 0.5$.

In the tunneling limit the ionization rate at a specific nuclear
separation is given as \cite{kansas2002}
\begin{align}
    \Gamma_{tun}(R)=&\sum_m\left(\frac{3F_0}{\pi\kappa^3}\right)^{\frac12}\nonumber
    \frac{B^2(m)}{2^{|m|}|m|!}\frac{1}{\kappa^{\frac{2Z}{\kappa-1}}}
    \left(\frac{2\kappa^3}{F_0}\right)^{\frac{2Z}{\kappa}-|m|-1}\\
&\times \exp\left({\frac{-2\kappa^3}{3F_0}}\right),
\end{align}
where $F_0$ is the electric field amplitude, $Z=2$ for H$_2^+$,
$\kappa =\sqrt{2 I_p(R)}$ and
\begin{align}
  B(m)=&\sum_lC_{l0}D_{m0}^l(R)(-1)^m\sqrt{\frac{(2l+1)(l+|m|)!}{2(l-|m|)!}}.
\end{align}
According to \cite{kansas2002} the effect of nuclear motion is
included by weighting the electronic ionization rate at different
internuclear separations with the probability of being at this
separation (the reflection principle);
\begin{align}
\frac{dw}{dR}=\Gamma_{tun}(R)|\chi_i(R)|^2.
\end{align}
This  result can be translated into a function of the proton kinetic
energy by assuming that all the Coulomb energy between the two
protons at the time of ionization is translated into proton ejection
energy. The results are shown in Fig.~3.

Compared to the strong-field approximation model the low energy part
of the spectrum shows better agreement with experiments, but the
high-energy part is again poorly represented. The reason for the
better agreement in the low-energy part of the spectrum is clear
from Figs.~3(b) and (d):  a shift towards lower vibrational states
and an exponential favoring of the low energy part of the spectrum
as is typical for a tunneling theory. The lack of agreement at high
proton energy could be an indication that  multiphoton
resonance ionization is involved at smaller bond lengths, and this
is clearly absent within the static field tunneling model.

Another way of using tunneling theory to describe Coulomb explosion
of H$_2^+$ is to include the effect of nuclear motion as described
in Ref.~\cite{Urbain2004}. The overall rate is here found by
\begin{align}
\frac{dw}{dE_{Nf}}=\left|\int\chi_f(R)\Gamma_{tun}^{\frac12}(R)\chi_i(R)dR\right|^2.
\end{align}
The results are given in Fig.~4, and show an even larger shift
towards lower vibrational excited states resulting in a slightly
better agreement particularly for the high intensity case.

\section{Conclusion}

In this work we have presented a relatively simple quantum
mechanical model to describe Coulomb explosion of molecular hydrogen
ions. The model builds on the length gauge molecular strong-field
approximation and gives reasonable predictions when compared with
experiment. Theory predicts a dominant proton emission at slightly
lower energies than the experimental measurements. However, the
effect of multiphoton resonance ionization contributions, known to
be significant for this range of intensities, could be important.

The strong-field approximation allows very simple and fast
calculations and might, for that reason, be a useful tool in the
further understanding of molecular dynamics, including related
intense field processes as, e.g., high-harmonic generation. For all
purposes it is important only to use the model in the regime, where
it is valid, namely describing pulses of high intensities and short
durations. If the pulse duration exceeds 50 fs the molecular ion
will have time to dissociate and the resulting KER-spectrum will
move to considerably lower energies.

For comparison we have discussed the predictions from two different
tunneling models. The spectrum is here somewhat narrower and also
centered near to low energies.

\begin{acknowledgments}
We thank Thomas Kim Kjeldsen for useful discussions. L.B.M. thanks
Xavier Urbain for discussions on the distribution over vibrational
levels and the University of Louvain-La-Neuve for hospitality and
support. The present work was supported by the Danish Research
Agency (Grant. No. 2117-05-0081).
\end{acknowledgments}

%\bibliography{bibfile}

\begin{thebibliography}{23}
\expandafter\ifx\csname natexlab\endcsname\relax\def\natexlab#1{#1}\fi
\expandafter\ifx\csname bibnamefont\endcsname\relax
  \def\bibnamefont#1{#1}\fi
\expandafter\ifx\csname bibfnamefont\endcsname\relax
  \def\bibfnamefont#1{#1}\fi
\expandafter\ifx\csname citenamefont\endcsname\relax
  \def\citenamefont#1{#1}\fi
\expandafter\ifx\csname url\endcsname\relax
  \def\url#1{\texttt{#1}}\fi
\expandafter\ifx\csname urlprefix\endcsname\relax\def\urlprefix{URL }\fi
\providecommand{\bibinfo}[2]{#2}
\providecommand{\eprint}[2][]{\url{#2}}

\bibitem[{\citenamefont{Giusti-Suzor et~al.}(1995)\citenamefont{Giusti-Suzor,
  Mies, DiMauro, Charron, and Yang}}]{review1995}
\bibinfo{author}{\bibfnamefont{A.}~\bibnamefont{Giusti-Suzor}},
  \bibinfo{author}{\bibfnamefont{F.~H.} \bibnamefont{Mies}},
  \bibinfo{author}{\bibfnamefont{L.~F.} \bibnamefont{DiMauro}},
  \bibinfo{author}{\bibfnamefont{E.}~\bibnamefont{Charron}}, \bibnamefont{and}
  \bibinfo{author}{\bibfnamefont{B.}~\bibnamefont{Yang}},
  \bibinfo{journal}{Journal of Physics B: Atomic, Molecular and Optical
  Physics} \textbf{\bibinfo{volume}{28}}, \bibinfo{pages}{309}
  (\bibinfo{year}{1995}),
  \urlprefix\url{http://stacks.iop.org/0953-4075/28/309}.

\bibitem[{\citenamefont{Posthumus}(2004)}]{belgien2004}
\bibinfo{author}{\bibfnamefont{J.~H.} \bibnamefont{Posthumus}},
  \bibinfo{journal}{Reports on Progress in Physics}
  \textbf{\bibinfo{volume}{67}}, \bibinfo{pages}{623} (\bibinfo{year}{2004}),
  \urlprefix\url{http://stacks.iop.org/0034-4885/67/623}.

\bibitem[{\citenamefont{McCann and Posthumus}(1999)}]{belfast1999}
\bibinfo{author}{\bibfnamefont{J.~F.} \bibnamefont{McCann}} \bibnamefont{and}
  \bibinfo{author}{\bibfnamefont{J.~H.} \bibnamefont{Posthumus}},
  \bibinfo{journal}{Philosophical Transactions of the Royal Society A:
  Mathematical, Physical and Engineering Sciences}
  \textbf{\bibinfo{volume}{357}}, \bibinfo{pages}{1309} (\bibinfo{year}{1999}),
  \urlprefix\url{http://dx.doi.org/10.1098/rsta.1999.0376}.

\bibitem[{\citenamefont{Ben-Itzhak
  et~al.}(2005{\natexlab{a}})\citenamefont{Ben-Itzhak, Wang, Xia, Sayler,
  Smith, Carnes, and Esry}}]{kansas2005}
\bibinfo{author}{\bibfnamefont{I.}~\bibnamefont{Ben-Itzhak}},
  \bibinfo{author}{\bibfnamefont{P.~Q.} \bibnamefont{Wang}},
  \bibinfo{author}{\bibfnamefont{J.~F.} \bibnamefont{Xia}},
  \bibinfo{author}{\bibfnamefont{A.~M.} \bibnamefont{Sayler}},
  \bibinfo{author}{\bibfnamefont{M.~A.} \bibnamefont{Smith}},
  \bibinfo{author}{\bibfnamefont{K.~D.} \bibnamefont{Carnes}},
  \bibnamefont{and} \bibinfo{author}{\bibfnamefont{B.~D.} \bibnamefont{Esry}},
  \bibinfo{journal}{Physical Review Letters} \textbf{\bibinfo{volume}{95}},
  \bibinfo{eid}{073002} (pages~\bibinfo{numpages}{4})
  (\bibinfo{year}{2005}{\natexlab{a}}),
  \urlprefix\url{http://link.aps.org/abstract/PRL/v95/e073002}.

\bibitem[{\citenamefont{Kulander et~al.}(1996)\citenamefont{Kulander, Mies, and
  Schafer}}]{california1996}
\bibinfo{author}{\bibfnamefont{K.~C.} \bibnamefont{Kulander}},
  \bibinfo{author}{\bibfnamefont{F.~H.} \bibnamefont{Mies}}, \bibnamefont{and}
  \bibinfo{author}{\bibfnamefont{K.~J.} \bibnamefont{Schafer}},
  \bibinfo{journal}{Phys. Rev. A} \textbf{\bibinfo{volume}{53}},
  \bibinfo{pages}{2562} (\bibinfo{year}{1996}).

\bibitem[{\citenamefont{Pavicic}(2004)}]{garching2004a}
\bibinfo{author}{\bibfnamefont{D.}~\bibnamefont{Pavicic}}, Ph.D. thesis,
  \bibinfo{school}{Max Planck Institute of Quantum Optics}
  (\bibinfo{year}{2004}).

\bibitem[{\citenamefont{Pavicic et~al.}(2003)\citenamefont{Pavicic, Kiess,
  Hänsch, and Figger}}]{garching2002}
\bibinfo{author}{\bibfnamefont{D.}~\bibnamefont{Pavicic}},
  \bibinfo{author}{\bibfnamefont{A.}~\bibnamefont{Kiess}},
  \bibinfo{author}{\bibfnamefont{T.~W.} \bibnamefont{Hänsch}},
  \bibnamefont{and} \bibinfo{author}{\bibfnamefont{H.}~\bibnamefont{Figger}},
  \bibinfo{journal}{The European Physical Journal D - Atomic, Molecular,
  Optical and Plasma Physics} \textbf{\bibinfo{volume}{26}},
  \bibinfo{pages}{39} (\bibinfo{year}{2003}),
  \urlprefix\url{http://dx.doi.org/10.1140/epjd/e2003-00197-2}.

\bibitem[{\citenamefont{Pavicic et~al.}(2005)\citenamefont{Pavicic, Kiess,
  Hansch, and Figger}}]{garching2004}
\bibinfo{author}{\bibfnamefont{D.}~\bibnamefont{Pavicic}},
  \bibinfo{author}{\bibfnamefont{A.}~\bibnamefont{Kiess}},
  \bibinfo{author}{\bibfnamefont{T.~W.} \bibnamefont{Hansch}},
  \bibnamefont{and} \bibinfo{author}{\bibfnamefont{H.}~\bibnamefont{Figger}},
  \bibinfo{journal}{Physical Review Letters} \textbf{\bibinfo{volume}{94}},
  \bibinfo{eid}{163002} (pages~\bibinfo{numpages}{4}) (\bibinfo{year}{2005}),
  \urlprefix\url{http://link.aps.org/abstract/PRL/v94/e163002}.

\bibitem[{\citenamefont{Zuo and Bandrauk}(1995)}]{canada1995}
\bibinfo{author}{\bibfnamefont{T.}~\bibnamefont{Zuo}} \bibnamefont{and}
  \bibinfo{author}{\bibfnamefont{A.~D.} \bibnamefont{Bandrauk}},
  \bibinfo{journal}{Phys. Rev. A} \textbf{\bibinfo{volume}{52}},
  \bibinfo{pages}{R2511} (\bibinfo{year}{1995}).

\bibitem[{\citenamefont{Peng et~al.}(2003)\citenamefont{Peng, Dundas, McCann,
  Taylor, and Williams}}]{belfast2003}
\bibinfo{author}{\bibfnamefont{L.-Y.} \bibnamefont{Peng}},
  \bibinfo{author}{\bibfnamefont{D.}~\bibnamefont{Dundas}},
  \bibinfo{author}{\bibfnamefont{J.~F.} \bibnamefont{McCann}},
  \bibinfo{author}{\bibfnamefont{K.~T.} \bibnamefont{Taylor}},
  \bibnamefont{and} \bibinfo{author}{\bibfnamefont{I.~D.}
  \bibnamefont{Williams}}, \bibinfo{journal}{Journal of Physics B: Atomic,
  Molecular and Optical Physics} \textbf{\bibinfo{volume}{36}},
  \bibinfo{pages}{L295} (\bibinfo{year}{2003}),
  \urlprefix\url{http://stacks.iop.org/0953-4075/36/L295}.

\bibitem[{\citenamefont{Ben-Itzhak
  et~al.}(2005{\natexlab{b}})\citenamefont{Ben-Itzhak, Wang, Xia, Sayler,
  Smith, Maseberg, Carnes, and Esry}}]{kansas2005a}
\bibinfo{author}{\bibfnamefont{I.}~\bibnamefont{Ben-Itzhak}},
  \bibinfo{author}{\bibfnamefont{P.}~\bibnamefont{Wang}},
  \bibinfo{author}{\bibfnamefont{J.}~\bibnamefont{Xia}},
  \bibinfo{author}{\bibfnamefont{A.~M.} \bibnamefont{Sayler}},
  \bibinfo{author}{\bibfnamefont{M.~A.} \bibnamefont{Smith}},
  \bibinfo{author}{\bibfnamefont{J.}~\bibnamefont{Maseberg}},
  \bibinfo{author}{\bibfnamefont{K.~D.} \bibnamefont{Carnes}},
  \bibnamefont{and} \bibinfo{author}{\bibfnamefont{B.~D.} \bibnamefont{Esry}},
  \bibinfo{journal}{Nuclear Instruments and Methods in Physics Research Section
  B: Beam Interactions with Materials and Atoms}
  \textbf{\bibinfo{volume}{233}}, \bibinfo{pages}{56}
  (\bibinfo{year}{2005}{\natexlab{b}}),
  \urlprefix\url{http://www.sciencedirect.com/science/article/B6TJN-4G1R3H9-2/%
2/e7dd0765b19273bea72ede330d2fefe4}.

\bibitem[{\citenamefont{Ergler et~al.}(2006)\citenamefont{Ergler, Rudenko,
  Feuerstein, Zrost, Schr\"{o}ter, Moshammer, and Ullrich}}]{garching2006}
\bibinfo{author}{\bibfnamefont{T.}~\bibnamefont{Ergler}},
  \bibinfo{author}{\bibfnamefont{A.}~\bibnamefont{Rudenko}},
  \bibinfo{author}{\bibfnamefont{B.}~\bibnamefont{Feuerstein}},
  \bibinfo{author}{\bibfnamefont{K.}~\bibnamefont{Zrost}},
  \bibinfo{author}{\bibfnamefont{C.~D.} \bibnamefont{Schr\"{o}ter}},
  \bibinfo{author}{\bibfnamefont{R.}~\bibnamefont{Moshammer}},
  \bibnamefont{and} \bibinfo{author}{\bibfnamefont{J.}~\bibnamefont{Ullrich}},
  \bibinfo{journal}{Journal of Physics B: Atomic, Molecular and Optical
  Physics} \textbf{\bibinfo{volume}{39}}, \bibinfo{pages}{S493}
  (\bibinfo{year}{2006}),
  \urlprefix\url{http://stacks.iop.org/0953-4075/39/S493}.

\bibitem[{\citenamefont{Saugout and Cornaggia}(2006)}]{frankrig2006}
\bibinfo{author}{\bibfnamefont{S.}~\bibnamefont{Saugout}} \bibnamefont{and}
  \bibinfo{author}{\bibfnamefont{C.}~\bibnamefont{Cornaggia}},
  \bibinfo{journal}{Physical Review A (Atomic, Molecular, and Optical Physics)}
  \textbf{\bibinfo{volume}{73}}, \bibinfo{eid}{041406}
  (pages~\bibinfo{numpages}{4}) (\bibinfo{year}{2006}),
  \urlprefix\url{http://link.aps.org/abstract/PRA/v73/e041406}.

\bibitem[{\citenamefont{McKenna et~al.}(2007)\citenamefont{McKenna, Bryan,
  Calvert, English, Wood, Murphy, Turcu, Smith, Ertel, Chekhlov
  et~al.}}]{belfast2007}
\bibinfo{author}{\bibfnamefont{J.}~\bibnamefont{McKenna}},
  \bibinfo{author}{\bibfnamefont{W.~A.} \bibnamefont{Bryan}},
  \bibinfo{author}{\bibfnamefont{C.~R.} \bibnamefont{Calvert}},
  \bibinfo{author}{\bibfnamefont{E.~M.~L.} \bibnamefont{English}},
  \bibinfo{author}{\bibfnamefont{J.}~\bibnamefont{Wood}},
  \bibinfo{author}{\bibfnamefont{D.~S.} \bibnamefont{Murphy}},
  \bibinfo{author}{\bibfnamefont{I.~C.~E.} \bibnamefont{Turcu}},
  \bibinfo{author}{\bibfnamefont{J.~M.} \bibnamefont{Smith}},
  \bibinfo{author}{\bibfnamefont{K.}~\bibnamefont{Ertel}},
  \bibinfo{author}{\bibfnamefont{O.}~\bibnamefont{Chekhlov}},
  \bibnamefont{et~al.}, \bibinfo{journal}{Journal of Modern Optics}
  \textbf{\bibinfo{volume}{in press}} (\bibinfo{year}{2007}).

\bibitem[{\citenamefont{Esry et~al.}(2006)\citenamefont{Esry, Sayler, Wang,
  Carnes, and Ben-Itzhak}}]{kansas2006}
\bibinfo{author}{\bibfnamefont{B.~D.} \bibnamefont{Esry}},
  \bibinfo{author}{\bibfnamefont{A.~M.} \bibnamefont{Sayler}},
  \bibinfo{author}{\bibfnamefont{P.~Q.} \bibnamefont{Wang}},
  \bibinfo{author}{\bibfnamefont{K.~D.} \bibnamefont{Carnes}},
  \bibnamefont{and}
  \bibinfo{author}{\bibfnamefont{I.}~\bibnamefont{Ben-Itzhak}},
  \bibinfo{journal}{Physical Review Letters} \textbf{\bibinfo{volume}{97}},
  \bibinfo{eid}{013003} (pages~\bibinfo{numpages}{4}) (\bibinfo{year}{2006}),
  \urlprefix\url{http://link.aps.org/abstract/PRL/v97/e013003}.

\bibitem[{\citenamefont{Kjeldsen and Madsen}(2004)}]{aarhus2004}
\bibinfo{author}{\bibfnamefont{T.~K.} \bibnamefont{Kjeldsen}} \bibnamefont{and}
  \bibinfo{author}{\bibfnamefont{L.~B.} \bibnamefont{Madsen}},
  \bibinfo{journal}{Journal of Physics B: Atomic, Molecular and Optical
  Physics} \textbf{\bibinfo{volume}{37}}, \bibinfo{pages}{2033}
  (\bibinfo{year}{2004}),
  \urlprefix\url{http://stacks.iop.org/0953-4075/37/2033}.

\bibitem[{\citenamefont{Kjeldsen and Madsen}(2005)}]{aarhus2005a}
\bibinfo{author}{\bibfnamefont{T.~K.} \bibnamefont{Kjeldsen}} \bibnamefont{and}
  \bibinfo{author}{\bibfnamefont{L.~B.} \bibnamefont{Madsen}},
  \bibinfo{journal}{Physical Review A (Atomic, Molecular, and Optical Physics)}
  \textbf{\bibinfo{volume}{71}}, \bibinfo{eid}{023411}
  (pages~\bibinfo{numpages}{10}) (\bibinfo{year}{2005}),
  \urlprefix\url{http://link.aps.org/abstract/PRA/v71/e023411}.

\bibitem[{\citenamefont{Peng et~al.}(2005)\citenamefont{Peng, Williams, and
  McCann}}]{belfast2005}
\bibinfo{author}{\bibfnamefont{L.-Y.} \bibnamefont{Peng}},
  \bibinfo{author}{\bibfnamefont{I.~D.} \bibnamefont{Williams}},
  \bibnamefont{and} \bibinfo{author}{\bibfnamefont{J.~F.}
  \bibnamefont{McCann}}, \bibinfo{journal}{Journal of Physics B: Atomic,
  Molecular and Optical Physics} \textbf{\bibinfo{volume}{38}},
  \bibinfo{pages}{1727} (\bibinfo{year}{2005}),
  \urlprefix\url{http://stacks.iop.org/0953-4075/38/1727}.

\bibitem{wigner} D.M. Brink and G.R. Satchler, {\it Angular
Momentum} (Oxford University Press, London, 1968); R.N. Zare, {\it
Angular Momentum} (Wiley, New York, 1988).

\bibitem[{\citenamefont{Gribakin and Kuchiev}(1997)}]{sydney1997}
\bibinfo{author}{\bibfnamefont{G.~F.} \bibnamefont{Gribakin}} \bibnamefont{and}
  \bibinfo{author}{\bibfnamefont{M.~Y.} \bibnamefont{Kuchiev}},
  \bibinfo{journal}{Phys. Rev. A} \textbf{\bibinfo{volume}{55}},
  \bibinfo{pages}{3760} (\bibinfo{year}{1997}).

\bibitem{aarhusextra} T.K. Kjelsden, C.Z. Bisgaard, L.B. Madsen, and H. Stapelfeldt,
Phys. Rev. A {\bf 71}, 013418 (2005).

\bibitem[{\citenamefont{Tong et~al.}(2002)\citenamefont{Tong, Zhao, and
  Lin}}]{kansas2002}
\bibinfo{author}{\bibfnamefont{X.~M.} \bibnamefont{Tong}},
  \bibinfo{author}{\bibfnamefont{Z.~X.} \bibnamefont{Zhao}}, \bibnamefont{and}
  \bibinfo{author}{\bibfnamefont{C.~D.} \bibnamefont{Lin}},
  \bibinfo{journal}{Phys. Rev. A} \textbf{\bibinfo{volume}{66}},
  \bibinfo{pages}{033402} (\bibinfo{year}{2002}).

\bibitem[{\citenamefont{Gradshteyn and Ryzhik}(1994)}]{Gradshteyn}
\bibinfo{author}{\bibfnamefont{I.~S.}~\bibnamefont{Gradshteyn}} \bibnamefont{and}
  \bibinfo{author}{\bibfnamefont{I.~M.} \bibnamefont{Ryzhik}},
  \emph{\bibinfo{title}{Table of Integrals, Series, and Products}} (\bibinfo{publisher}{Academic Press, San Diego}, \bibinfo{year}{1994}).

\bibitem[{\citenamefont{Niikura et~al.}(2002)\citenamefont{Niikura, Légaré,
  Hasbani, Bandrauk, Ivanov, Villeneuve, and Corkum}}]{canada2002}
\bibinfo{author}{\bibfnamefont{H.}~\bibnamefont{Niikura}},
  \bibinfo{author}{\bibfnamefont{F.}~\bibnamefont{Légaré}},
  \bibinfo{author}{\bibfnamefont{R.}~\bibnamefont{Hasbani}},
  \bibinfo{author}{\bibfnamefont{A.~D.} \bibnamefont{Bandrauk}},
  \bibinfo{author}{\bibfnamefont{M.~Y.} \bibnamefont{Ivanov}},
  \bibinfo{author}{\bibfnamefont{D.~M.} \bibnamefont{Villeneuve}},
  \bibnamefont{and} \bibinfo{author}{\bibfnamefont{P.~B.}
  \bibnamefont{Corkum}}, \bibinfo{journal}{Nature}
  \textbf{\bibinfo{volume}{417}}, \bibinfo{pages}{917} (\bibinfo{year}{2002}).

\bibitem[{\citenamefont{Urbain et~al.}(2004)\citenamefont{Urbain, Fabre,
  Staicu-Casagrande, de~Ruette, Andrianarijaona, Jureta, Posthumus, Saenz,
  Baldit, and Cornaggia}}]{Urbain2004}
\bibinfo{author}{\bibfnamefont{X.}~\bibnamefont{Urbain}},
  \bibinfo{author}{\bibfnamefont{B.}~\bibnamefont{Fabre}},
  \bibinfo{author}{\bibfnamefont{E.~M.} \bibnamefont{Staicu-Casagrande}},
  \bibinfo{author}{\bibfnamefont{N.}~\bibnamefont{de~Ruette}},
  \bibinfo{author}{\bibfnamefont{V.~M.} \bibnamefont{Andrianarijaona}},
  \bibinfo{author}{\bibfnamefont{J.}~\bibnamefont{Jureta}},
  \bibinfo{author}{\bibfnamefont{J.~H.} \bibnamefont{Posthumus}},
  \bibinfo{author}{\bibfnamefont{A.}~\bibnamefont{Saenz}},
  \bibinfo{author}{\bibfnamefont{E.}~\bibnamefont{Baldit}}, \bibnamefont{and}
  \bibinfo{author}{\bibfnamefont{C.}~\bibnamefont{Cornaggia}},
  \bibinfo{journal}{Physical Review Letters} \textbf{\bibinfo{volume}{92}},
  \bibinfo{eid}{163004} (pages~\bibinfo{numpages}{4}) (\bibinfo{year}{2004}),
  \urlprefix\url{http://link.aps.org/abstract/PRL/v92/e163004}.

\end{thebibliography}

\end{document}